\documentclass[reqno,11pt]{article}

\RequirePackage[OT1]{fontenc}
\RequirePackage{amsthm, amsmath, amssymb, amsfonts, natbib, xspace, graphicx, enumerate,multirow}

\renewcommand{\baselinestretch}{1.5}

\theoremstyle{plain}
\newtheorem{theorem}{Theorem}[section]
\newtheorem{lemma}{Lemma}[section]
\newtheorem{corollary}{Corollary}[section]

\theoremstyle{definition}

\theoremstyle{remark}

\newcommand{\eq}[1]{(\ref{eq:#1})\xspace}

\newcommand{\sbul}{\lower 3pt\hbox{\LARGE $\cdot\,$}}

\newcommand{\bA}{{\mbox{\boldmath $A$}}}
\newcommand{\bH}{{\mbox{\boldmath $H$}}}
\newcommand{\bI}{{\mbox{\boldmath $I$}}}

\newcommand{\bv}{{\mbox{\boldmath $v$}}}
\newcommand{\bX}{{\mbox{\boldmath $X$}}}
\newcommand{\bx}{{\mbox{\boldmath $x$}}}
\newcommand{\bY}{{\mbox{\boldmath $Y$}}}
\newcommand{\by}{{\mbox{\boldmath $y$}}}
\newcommand{\byi}{{\by_\bi}}
\newcommand{\byic}{{\by_\bi^\complement}}
\newcommand{\bu}{{\mbox{\boldmath $u$}}}
\newcommand{\bU}{{\mbox{\boldmath $U$}}}

\newcommand{\bZ}{{\mbox{\boldmath $Z$}}}
\newcommand{\bd}{{\mbox{\boldmath $d$}}}
\newcommand{\bG}{{\mbox{\boldmath $G$}}}
\newcommand{\bi}{{\mbox{\boldmath $i$}}}
\newcommand{\bic}{{\bi^\complement}}

\newcommand{\iid}{i.i.d.\xspace}

\newcommand{\bmu}{{\mbox{\boldmath $\mu$}}}
\newcommand{\bepsilon}{{\mbox{\boldmath $\epsilon$}}}
\newcommand{\bbeta}{{\mbox{\boldmath $\beta$}}}
\newcommand{\btheta}{{\mbox{\boldmath $\theta$}}}
\newcommand{\bTheta}{{\mbox{\boldmath $\Theta$}}}
\newcommand{\fidu}{proposed}

\newcommand{\ignore}[1]{}

\begin{document}
\title{\vspace*{-2cm}Generalized Fiducial Inference for Ultrahigh Dimensional Regression
\footnote{The authors thank Professors Jianqing Fan and Ning Hao for sharing the housing price appreciation data set.  The work of Hannig was supported in part by the National Science Foundation under Grants 1007543 and 1016441.  The work of Lee was supported in part by the National Science Foundation under Grants 1007520, 1209226 and 1209232.}
}

\author{
Randy C. S. Lai\thanks{Department of Statistics, University of California at Davis, 4118 Mathematical Sciences Building, One Shields Avenue, Davis, CA 95616, USA. Email: {\ttfamily rcslai@ucdavis.edu}} \\
University of California at Davis
\and
Jan Hannig\thanks{Department of Statistics and Operations Research, University of North Carolina at Chapel Hill, Chapel Hill, NC 27599-3260, USA.  Email: {\ttfamily jan.hannig@unc.edu}}\\
{University of North Carolina at Chapel Hill}
\and
Thomas C. M. Lee\thanks{Corresponding authour.  Department of Statistics, University of California at Davis, 4118 Mathematical Sciences Building, One Shields Avenue, Davis, CA 95616, USA. Email: {\ttfamily tcmlee@ucdavis.edu}} \\
%Phone: +1(530) 752 2361; Fax: +1 (530) 752 7099}\\
University of California at Davis
}

\author{
Randy C. S. Lai\thanks{Department of Statistics, University of California at Davis, 4118 Mathematical Sciences Building, One Shields Avenue, Davis, CA 95616, USA. Email: {\ttfamily rcslai@ucdavis.edu}}
\and
Jan Hannig\thanks{Department of Statistics and Operations Research, University of North Carolina at Chapel Hill, Chapel Hill, NC 27599-3260, USA.  Email: {\ttfamily jan.hannig@unc.edu}}
\and
Thomas C. M. Lee\thanks{Corresponding author.  Department of Statistics, University of California at Davis, 4118 Mathematical Sciences Building, One Shields Avenue, Davis, CA 95616, USA. Email: {\ttfamily tcmlee@ucdavis.edu}} \\
%Phone: +1(530) 752 2361; Fax: +1 (530) 752 7099}
}

\date{April 29, 2013}
%\author{}

\maketitle

\begin{abstract}

In recent years the ultrahigh dimensional linear regression problem has attracted enormous attentions from the research community.  Under the sparsity assumption most of the published work is devoted to the selection and estimation of the significant predictor variables.  This paper studies a different but fundamentally important aspect of this problem: uncertainty quantification for parameter estimates and model choices.  To be more specific, this paper proposes methods for deriving a probability density function on the set of all possible models, and also for constructing confidence intervals for the corresponding parameters.  These proposed methods are developed using the generalized fiducial methodology, which is a variant of Fisher's controversial fiducial idea.  Theoretical properties of the proposed methods are studied, and in particular it is shown that statistical inference based on the proposed methods will have exact asymptotic frequentist property.  In terms of empirical performances, the proposed methods are tested by simulation experiments and an application to a real data set.  Lastly this work can also be seen as an interesting and successful application of Fisher's fiducial idea to an important and contemporary problem.  To the best of the authors' knowledge, this is the first time that the fiducial idea is being applied to a so-called ``large $p$ small $n$'' problem.

\end{abstract}
%\vfill
%\pagebreak

\section{Introduction}

The ultrahigh dimensional linear regression problem has attracted enormous attentions in recent years.  A typical description of the problem begins with the usual linear model
\[
Y_i=\bx_i^T\bbeta + \epsilon_i, \quad \mbox{or equivalently} \quad \bY=\bX\bbeta+\bepsilon,
\]
where $\bY=(Y_1, \ldots, Y_n)^T$ is a vector of $n$ responses, $\bX=(\bx_1, \ldots, \bx_n)^T$ is a design matrix of size $n\times p$ with \iid variables $\bx_1, \ldots, \bx_n$, $\bbeta=(\beta_1, \ldots, \beta_p)^T$ is a vector of $p$ parameters, and $\bepsilon=(\epsilon_1, \ldots, \epsilon_n)^T$ is a vector of $n$ \iid random errors with zero mean and unknown variance $\sigma^2$.  It is assumed that $\bepsilon$ and $\bx_1, \ldots, \bx_n$ are independent, and that $p$ is larger than $n$ and grows at an exponential rate as $n$ increases.  It is this last assumption that makes the ultrahigh dimensional regression problem different from the classical multiple regression problem, for which $p<n$.

When $p\gg n$, it is customary to assume that the number of significant predictors in the true model is small; i.e., the true model is sparse.  The problem is then to identify which $\beta_j$'s are non-zero, and to estimate their values.  To solve this variable selection problem, one common strategy is to first apply a so-called screening procedure to remove a large number of insignificant predictors, and then apply a penalized method such as the LASSO method of \citet{tibshirani1996regression} or the SCAD method of \citet{Fan-Li01} to the surviving predictors to select the final set of variables.  For screening procedures, one of the earliest is the sure independence screening procedure of \citet{fan2008sure}.  Since then various screening procedures have been proposed: \citet{wang2009forward} developed a consistent screening procedure that combines forward regression and the extended BIC criterion of \citet{ChenChen2008}, \citet{buhlmann2010variable} proposed a screening procedure that is based on conditional partial corrections, and \citet{cho2011high} constructed a screening procedure that utilizes information from both marginal correlation and tilted correlation.  Also, other screening procedures are developed for more complicated settings, including generalized linear models and nonparametric additive modeling; e.g., \citet{meier2009high}, \citet{ravikumar2009sparse}, \citet{fan2011non-concave} and \citet{fan2011nonparametric}.  For an overview of variable selection for high dimensional problems, see \citet{fan2010selective}.

While much efforts have been spent on model selection and parameter estimation for the ultrahigh dimensional regression problem, virtually no published work is devoted to quantify the uncertainty in the chosen models and their parameter estimates.  A notable exception is the pioneer work of \citet{Fan2011}, where a cross-validation based method is proposed to estimate the error variance $\sigma^2$.  Given such an estimate and a final model, confidence intervals for $\beta_j$'s can be constructed using classical linear model theory.  However, this approach does not account for the additional variability contributed by the need of selecting a final model.

The goal of this paper is to investigate the use of Fisher's fiducial idea \citep{Fisher1930} in the ultrahigh dimensional regression problem.  In particular a new procedure is developed for constructing confidence intervals for all the parameters (including $\sigma$) in the final selected model.  This procedure automatically accounts for the variability introduced by model selection.  To the best of our knowledge, this is the first time that Fisher's fiducial idea is being applied to the so-called ``large $p$ small $n$'' problem.

\citet{Fisher1930} introduced fiducial inference in order to define a statistically meaningful distribution on the parameter space in cases when one cannot use a Bayes theorem due to the lack of prior information. While never formally defined, fiducial inference has a long and storied history. We refer an interested reader to \citet{Hannig2009} and \citet{Salome1998} where a wealth of references can be found.

Ideas related to fiducial inference has experienced an exciting resurgence in the last decade.  Some of these modern ideas are Dempster-Shafer calculus and its generalizations \citep{Dempster2008,MartinZhangLiu2010,ZhangLiu2011, martin2013inferential}, confidence distributions \citep{SinghXieStrawderman2005,XieSinghStrawderman2011}, generalized inference \citep{Weerahandi1993,Weerahandi1995} and reference priors in objective Bayesian inference \citep{BergerBernardoSun2009}. There has also been a wealth of successful applications of these methods to practical problems. For selected examples see \citet{
%KrishnamoorthyLinXia2009,
%LiuLiuXie2011,
McNallyIyerMathew2003,
WangIyer2005,
%WangIyer2006a,
%WangIyer2006b,
%WangIyer2008,
EHannigIyer2008,
EdlefsenLiuDempster2009,
HannigLee2009}
%WandlerHannig2012a,
%WandlerHannig2012b,
%WandlerHannig2011
and \citet{CisewskiHannig2012}.

The particular variant of Fisher's fiducial idea that this paper considers is the so-called {\em generalized fiducial inference}.  Some early ideas were developed by \cite{HannigIyerPatterson2006}, and later \cite{Hannig2009} used these ideas to formally define a {\em generalized fiducial distribution}.  An brief description of generalized fiducial inference is given below.

The rest of this paper is organized as follows.  Section~\ref{sec:main} provides some background material on generalized fiducial inference, and applies the methodology to the ultrahigh dimensional regression problem.  The theoretical properties of the proposed solution are examined in Section~\ref{sec:theory}, while its empirical properties are illustrated in Section~\ref{sec:empirical}.  Lastly, concluding remarks are offered in Section~\ref{sec:conclude} and technical details are delayed to the appendix.

%\section{Generalized Fiducial Inference for Ultrahigh Dimensional Regression}
\section{Methodology}
\label{sec:main}
Generalized fiducial inference begins with expressing the relationship between the data $\bY$ and the parameters $\btheta$ as
\begin{equation}
\bY=\bG(\bU, \btheta),
\label{eqn:structural}
\end{equation}
where $\bG(\cdot,\cdot)$ is sometimes known as the {\em structural equation}, and $\bU$ is the random component of the relation whose distribution is {\em completely known}; e.g., a vector of \iid U(0,1)'s.  Recall that in the definition of the celebrated maximum likelihood estiimator, Fisher ``switched'' the roles of $\bY$ and $\btheta$: the random $\bY$ is treated as deterministic in the likelihood function, while the deterministic $\btheta$ is treated as random.  Through~(\ref{eqn:structural}) generalized fiducial inference uses this ``switching principle'' to define a valid probability distribution on $\btheta$.

This switching principle proceeds as follows.  For the moment suppose for any given realization $\by$ of $\bY$, the inverse
\begin{equation}
\btheta=\tilde{\bG}^{-1}(\by,\bu)
\label{eqn:inverse}
\end{equation}
always exists for any realization $\bu$ of $\bU$.  Since the distribution of $\bU$ is assumed known, one can always generate a random sample $\tilde{\bu}_1, \tilde{\bu}_2, \ldots$, and via~(\ref{eqn:inverse}) a random sample of $\btheta$ can be obtained by $\tilde{\btheta}_1=\tilde{\bG}^{-1}(\by,\tilde{\bu}_1), \tilde{\btheta}_2=\tilde{\bG}^{-1}(\by,\tilde{\bu}_2), \ldots$.  This is called a fiducial sample of $\btheta$, which can be used to calculate estimates and construct confidence intervals for $\btheta$ in a similar fashion as with a Bayesian posterior sample.  Through the above switching and the inverse operations, one can see that a density function $r(\btheta)$ for $\btheta$ is implicitly defined.  We term $r(\btheta)$ the {\em generalized fiducial density} for $\btheta$, and the corresponding distribution the {\em generalized fiducial distribution} for $\btheta$.  An illustrative example of applying this idea to simple linear regression can be found in \cite{HannigLee2009}, and a formal mathematical definition of generalized fiducial inference is described in detail in \cite{Hannig2009}.  The latter work also provides strategies to ensure the existence of the inverse~(\ref{eqn:inverse}).

Observe that for the ultrahigh dimensional regression problem that this paper considers, $\btheta$ can be decomposed into three components: $\btheta=\{M, \sigma, \bbeta_M\}$, where $M$ denotes a candidate model and can be seen as a sequence of $p$ binary variables indicating which predictors are significant, $\sigma$ is the noise standard deviation and $\bbeta_M$ is the coefficients of the significant predictors.  In the next subsection we derive the generalized fiducial density $r(M)$ for $M$, and then we will demonstrate how to generate a fiducial sample $\{\tilde{M}, \tilde{\sigma}, \tilde{\bbeta}\}$ using $r(M)$.

\subsection{Generalized Fiducial Density for Ultrahigh Dimensional Regression}
While the above formal definition of generalized fiducial inference is conceptually simple and very general, it may not be easily applicable in some practical situations.  When the model dimension is known, \cite{Hannig2012} derived a workable formula for $r(\btheta)$ for many practical situations.  Assume that the parameter $\btheta\in\bTheta\subset\mathbb R^d$ is $d$-dimensional and that the inverse $\bG^{-1}(\by,\btheta) = \bu$ to (\ref{eqn:structural}) exists.  This assumption is satisfied for many natural structural equations, provided that $\by$ and $\bu$ have the same dimension and $\bG$ is smooth.  Note that this inverse is different from the inverse $\tilde{\bG}^{-1}$ in~(\ref{eqn:inverse}).  Then under some differentiability assumptions, \citet{Hannig2012} showed that the generalized fiducial distribution is absolutely continuous with density
\begin{equation}\label{eq:GreatFiducial}
    r(\btheta)=\frac{f(\by,\btheta) J(\by,\btheta)}{\int_\bTheta f(\by,\btheta') J(\by,\btheta')\,d\btheta'},
\end{equation}
where
\begin{equation}\label{eq:RecommendedJacobian1}
 J(\by,\btheta)= \sum_{\substack{\bi=(i_1,\ldots,i_d) \\ 1\leq i_1<\cdots<i_d\leq n}}\left|\det\left[\left\{\frac{\bd}{\bd\by} \bG^{-1}(\by,\btheta)\right\}^{-1}\frac{\bd}{\bd(\btheta, \byic)} \bG^{-1}(\by,\btheta)\right]\right|.
\end{equation}
In the above $f(\by,\btheta)$ is the likelihood and the sum goes over all $p$-tuples of indexes $\bi=(1\leq i_1<\cdots< i_d\leq n)\subset\{1, \ldots, n\}$.  Also, for each $\bi$ we denoted the list of unused indexes by $\bic=\{1,\ldots,n\}\setminus\bi$, the collection of variables indexed by $\bi$ by $\byi=(y_{i_1},\ldots,y_{i_d})$, and its complement by $\byic=(y_i\,:\, i\in\bic)$.  The formula $\frac{\bd}{\bd(\btheta, \byic)} \bG^{-1}(\by,\btheta)$ stood for the Jacobian matrix computed with respect to all parameters $\btheta$ and the observations $\by_\bic$. Similarly $\frac{\bd}{\bd\by} \bG^{-1}(\by,\btheta)$ stood for the Jacobian matrix computed with respect to the observations $\by$.

Recall that the formula~(\ref{eq:GreatFiducial}) was derived for situations where the model dimension is known, and hence it cannot be directly applied to the current problem.  When model selection is required, \cite{HannigLee2009} proposed adding extra penalty structural equations to~(\ref{eq:GreatFiducial}).  This is similar to adding a penalty term to the likelihood function to account for model complexity.  In particular their derivation shows that the fiducial probability of each candidate model $M$ is proportional to
\begin{equation}\label{eq:penalizedfiducia}
  r(M)\propto {\int_\bTheta f_M(\by,\btheta) J_M(\by,\btheta)\,d\btheta} \, e^{-q(M)},
\end{equation}
where $f_M(\by,\btheta)$ is the likelihood, $J_M(\by,\btheta)$ is the Jacobian~(\ref{eq:RecommendedJacobian1}), and $q(M)$ is the penalty associated with the model $M$.  In the context of wavelet regression, they recommended using the minimum description length (MDL) principle \citep{Rissanen89, Rissanen07} to derive the penalty $q(M)$, which is shown to possess attractive theoretical and empirical properties.

Given the success of \cite{HannigLee2009}, we also attempted to use the MDL principle to derive a penalty $q(M)$ for the current problem, which gives $q(M)=0.5|M|\log n$ with $|M|$ being the number of significant parameters in $M$.  However, this form of $q(M)$ fails here, as the classical MDL principle was not designed to handle the ``$p\gg n$'' scenario.  To overcome this issue, we propose using the following penalty
\begin{equation}\label{eq:MDLpenalty}
 q(M)=\frac{|M|}{2}\log n + \log_{e^{1/\gamma}} \binom{p}{|M|},
\end{equation}
where the additional second term comes from the need to encode which of the parameters are left as zero.  Here $\gamma$ is a constant measuring the quality of the encoding; the most natural choice is $\gamma=1$ but other choices are possible.  In all our numerical work we use $\gamma=1$.  We note that the second term of~(\ref{eq:MDLpenalty}) is similar to the EBIC penalty of \cite{ChenChen2008}.

Denote the residual sum of squares of $M$ as $\text{RSS}_M$ when the corresponding $\bbeta$ is estimated with maximum likelihood.  Using penalty~(\ref{eq:MDLpenalty}), for the current ultrahigh dimensional regression problem, it is shown in Appendix~\ref{app:derivegfd} that the fiducial probability for model $M$ is
\begin{equation}\label{eq:FidModelProb}
r(M)\propto  \Gamma\left( \frac{n-|M|}{2} \right) \left(\pi \text{RSS}_{M}\right)^{-\frac{n-|M|-1}{2}} n^{-\frac{|M|+1}2} \binom{p}{|M|}^{-\gamma}.
\end{equation}

\subsection{Practical Generation of Fiducial Sample}
\label{sec:pa}
In this subsection we propose a practical procedure for generating a fiducial sample $\{\tilde{M},\tilde{\sigma},\tilde{\bbeta}\}$ using~(\ref{eq:FidModelProb}).  First note that even for a moderate $p$, the total number of models $2^p$ is huge and hence any method that is exhaustive in nature is computationally not feasible.

The proposed procedure begins with constructing a class of candidate models, denoted as $\mathcal M'$.  This $\mathcal M'$ should satisfy the following two properties: $|\mathcal M'|$ is small and it contains the true model and models that have non-negligible values of $r(M)$.  To construct $\mathcal M'$, we first apply the sure independence screening (SIS) procedure of \citet{fan2008sure} to reduce the number of predictors from $p$ to $p'$, where $p'$ is of order $O(n)$.  To further reduce the number of possible models (which is $2^{p'}$), we apply LASSO to those $p'$ predictors that survived SIS, and take all those models that lie on the LASSO solution path as $\mathcal M'$.  Note that the LASSO solution path can be quickly obtained via the least angle regression method \citep{efron2004least}, and that constructing $\mathcal M'$ in this way will ensure the true model is captured in $\mathcal M'$ with probability 1 \citep{fan2008sure}.

Once $\mathcal M'$ is obtained, for each $M \in \mathcal M'$, calculate
\[
R(M)=\Gamma\left( \frac{n-|M|}{2} \right) \left(\pi \text{RSS}_{M}\right)^{-\frac{n-|M|-1}{2}} n^{-\frac{|M|+1}2} \binom{p}{|M|}^{-\gamma},
\]
and approximate the generalized fiducial probability~(\ref{eq:FidModelProb}) by
\begin{align}
\label{eq:fdist1}
r(M) \approx R(M) / \sum_{M' \in \mathcal M'} R(M'), \quad \text{for } M \in \mathcal M'.
\end{align}

Next for $\sigma$ and $\bbeta_M$.  For any given $M$, it is straightforward to show that the generalized fiducial distribution of $\sigma$ conditional on $M$ is
\begin{align}
\label{eq:fdist2}
\text{RSS}_M/ \sigma^2 \sim \chi^2(n-|M|)
\end{align}
and that of $\bbeta_M$ conditional on $M$ and $\sigma$ is
\begin{align}
\label{eq:fdist3}
\bbeta_M \sim N(\bbeta_M^{\rm ML}, \sigma^2 \bX_M^T \bX_M),
\end{align}
where $\bbeta_M^{\rm ML}$ is the maximum likelihood estimate of $\bbeta_M$ for model $M$, and $\bX_M$ is the design matrix for model $M$.

Thus to generate $\{\tilde{M}, \tilde{\sigma}, \tilde{\bbeta}\}$, we first draw a model $\tilde{M} \in {\mathcal M'}$ from \eq{fdist1}, then $\tilde{\sigma}$ from \eq{fdist2} given $\tilde{M}$, and lastly $\tilde{\bbeta}$ from \eq{fdist3} given $\{\tilde{M}, \tilde{\sigma}\}$.

\subsection{Point Estimates and Confidence Intervals}
Applying the above procedure repeatedly one can obtain multiple copies of $\{\tilde{M}, \tilde{\sigma}, \tilde{\bbeta}\}$ that form a fiducial sample for $\{M, \sigma, \bbeta_M\}$.  This fiducial sample can be used to form estimates and confidence intervals for $\sigma$ in a similar manner as with a Bayesian posterior sample.  For example, the average of all $\tilde{\sigma}$'s can be used as an estmate of $\sigma$, while the 2.5\% smallest and 2.5\% largest $\tilde{\sigma}$ values can be used respectively as the lower and upper limits for a 95\% confidence interval for $\sigma$.

Obtaining estimates and confidence intervals for $\bbeta$ is, however, less straightforward.  It is because for any $\beta_j$, it is possible that it is included in some but not all $\tilde{M}$'s.  In other words, some of the generated fiducial values for $\beta_j$ are zeros, some are not.

We use the following simple procedure to deal with this issue.  For each $\beta_j$, we count the percentage of zero fiducial sample values.  If it is more than 50\%, we declare that this particular $\beta_j$ is not significant.  Otherwise, we treat $\beta_j$ as a significant parameter, and use all the non-zero fiducial sample values to obtain estimates and confidence intervals for it, in the same way as for $\sigma$.  Note that a similar idea has been used by \citet{barbieri2004optimal} to determine the significance of a parameter in the Bayesian context.

\section{Theoretical Properties}
\label{sec:theory}
This section investigates the theoretical properties of the above-proposed generalized fiducial based method, under the situation that $p$ is diverging and the size of true model is either fixed or diverging.  For similar results in the classical situations where $p$ is fixed, see \cite{Hannig2009, Hannig2012}.

First, some notations.  Let $M$ be any model, $M_0$ be the true model, and $\bH_M$ be the projection matrix of $\bX_M$; i.e., $\bH_M = \bX_M(\bX_M^T \bX_M)^{-1} \bX_M^T$.  Define
\[
\Delta_M = ||\bmu - \bH_M \bmu||^2,
\]
where $\bmu = E (\bY) = \bX_{M_0} \bbeta_{M_0}$.  Throughout this section we assume the following identifiability condition holds:
\begin{align}
    \lim_{n\to \infty} \min \left\{ \frac{\Delta_M}{|M_0| \log p}: \ M_0 \not\subset M, |M| \le k |M_0| \right\} = \infty \label{eq:idCon}
\end{align}
for some fixed $k>1$.  This condition ensures that the true model is identifiable and has been used for example by \cite{luo2011}.  It can be shown that, under the sparse Reisz condition and the condition
\[
\sqrt{\frac{n}{|M_0| \log p}} \min\left\{ |\beta_j|; j \in M_0 \right\} \to \infty,
\]
the identifiability condition \eq{idCon} holds.  However, the inverse does not hold in general.

Let $\mathcal M$ be the collection of models such that $\mathcal M = \left\{ M : \ |M| \le k |M_0| \right\}$ for some fixed $k$.  The restriction $|M| \le k |M_0|$ is imposed because in practice we only consider models with size comparable with the true model.

If $p$ is large, the size of $\mathcal M$ could still be too large in practice.  In this situation, we could use a variable screening procedure to reduce the size.  This variable screening procedure should result in a class of candidate models $\mathcal M'$ which satisfies
\begin{align}
P(M_0 \in \mathcal M') \to 1 \quad \text{and} \quad \log(|\mathcal M'_j|) = o(j \log n),
\label{eq:selectCon}
\end{align}
where $\mathcal M'_j$ contains all models in $\mathcal M'$ that are of size $j$.  The first condition in \eq{selectCon} guarantees the model class contains the true model, at least asymptotically. The second condition in \eq{selectCon} ensures that the size of the model class is not too large.  These two conditions are satisfied by the practical algorithm presented in Section~\ref{sec:pa}.

In Appendix~\ref{app:proofs} the following theorem is established.

\begin{theorem}
    \label{th:main}
     Under \eq{idCon}, as $n \to \infty$, $p \to \infty$, $|M_0| \log(p) = o(n)$, $\log (|M_0|) / \log(p) \to \delta$ and $\log(n)/\log(p) \to \eta$, then there exists $\gamma> \frac{1+\delta}{1-\delta} - \frac{3\eta}{2(1-\delta)}$ such that
    \begin{align}
        \max_{M \ne M_0, M \in \mathcal M} r(M)/r(M_0) \overset{P}{\to} 0.  \label{eq:thm1}
    \end{align}
Furthermore, if \eq{selectCon} holds, with the same $\gamma$,
    \begin{align}
r(M) \overset{P}{\to}  1 \label{eq:thm1.2}
  \end{align}
  over the class $\mathcal M'$.
\end{theorem}
Equation \eq{thm1} states that the true model has the highest generalized fiducial probability amongst all the models in $\mathcal M$.  However, it does not imply equation \eq{thm1.2} in general because the class of candidate models can be very large.  If we constrain the class of models being considered in such a way that \eq{selectCon} holds, then equation~\eq{thm1.2} states that, with probability tending to 1, the true model will be selected.  From Theorem~\ref{th:main}, one can conclude the following important corollary.
\begin{corollary}
    Statistical inference that is based on the generalized fiducial density~\eq{FidModelProb} will have exact asymptotic frequentist property. Consequently the generalized fiducial distribution and derived point estimators are consistent.
\end{corollary}

\section{Finite Sample Properties}
\label{sec:empirical}
\subsection{Simulations}
A simulation study was conducted to evaluate the practical performance of the proposed procedure.  The following model from \citet{Fan2011} was used to generate the noisy data
\[
Y = b (X_1 +\cdots + X_d) + \epsilon,
\]
where $\epsilon$ is \iid standard normal error, $d$ is the number of significant predictors, and the coefficient $b$ controls the signal-to-noise ratio.  All the covariates are standard normal variables with correlation $\mbox{cor}(X_i,X_j) = \rho^{|i-j|}$.  Three combinations of $(n, p, d)$ were used: $(200, 2000, 3)$, $(300, 8000, 5)$ and $(500, 50000, 8)$.  For each of these three combinations, 3 choices of $b$ and 2 choices of $\rho$ were used: $b=1/\sqrt{d}, 2/\sqrt{d}$ and $3/\sqrt{d}$, and $\rho=0$ and $0.5$.  Therefore, a total of $3\times 3\times 2 = 18$ experimental configurations were considered.  The number of repetitions for each experimental configuration was 1000.  For $\rho=0$, the cases $b = 1/\sqrt{d}, 2/\sqrt{d}$ and $3/\sqrt{d}$ correspond to the cases when the signal-to-noise ratios are 1, 2 and 3 respectively.

For each generated data set, we applied the proposed generalized fiducial procedure described in Section~\ref{sec:pa} to obtain a fiducial sample of size 10,000 for $\{M, \sigma, \bbeta\}$, and from this we computed the generalized fiducial estimate for $\sigma^2$.  We also obtained two other estimates for $\sigma^2$: the first one from the refitted cross-validation (RCV) method of \citet{Fan2011}, while the second one is the classical maximum likelihood estimate for $\sigma^2$ obtained from the true model.  Of course the last estimate cannot be obtained in practice, but it is computed here for benchmark comparisons.  In sequel it is termed as the oracle estimate.  Also, for RCV, the particular version we compared with is RCV-LASSO.

The bias of these three estimates for $\sigma^2$ are summarized in Table~\ref{table:sigmaestimate}.  From this table one can see that the bias of the fiducial estimates are usually not much larger than the bias from the oracle estimates.  The RCV estimates sometimes have very large bias.

\begin{table}[ht]
    \centering
    \renewcommand{\baselinestretch}{1.2}
    \scriptsize
    \begin{tabular}{|c|c|ccc|}\hline
        \multicolumn{2}{|c|}{}  & $(n,p,d)=(200,2000,3)$ & $(n,p,d)=(300,8000,5)$ &$(n,p,d)=(500,50000,8)$\\\hline
        \multirow{3}{*}{\begin{minipage}{2cm} $b=1/\sqrt{3}$ \\$\rho=0$ \end{minipage}}
& \fidu & $-0.180$ (0.323) & $-0.166$ (0.271) & 0.230 (0.219) \\
& RCV& 1.507 (0.488) & $-16.749$ (0.330) & $-27.287$ (0.221) \\
%& EBIC& -0.513 (0.321) & -0.274 (0.275) & 0.181 (0.225) \\
& oracle& $-0.018$ (0.317) & $-0.115$ (0.263) & $-0.031$ (0.200) \\\hline
        \multirow{3}{*}{\begin{minipage}{2cm} $b=2/\sqrt{3}$ \\$\rho=0$ \end{minipage}}
& \fidu & $-0.511$ (0.327) & $-0.455$ (0.259) & $-0.089$ (0.202) \\
& RCV& $-0.297$ (0.465) & $-7.932$ (0.353) & $-13.909$ (0.255) \\
%& EBIC& -0.769 (0.329) & -0.632 (0.26) & -0.247 (0.201) \\
& oracle& $-0.383$ (0.321) & $-0.474$ (0.260) & $-0.151$ (0.200) \\\hline
        \multirow{3}{*}{\begin{minipage}{2cm} $b=3/\sqrt{3}$ \\$\rho=0$ \end{minipage}}
& \fidu & $-0.457$ (0.332) & $-0.112$ (0.256) & 0.103 (0.203) \\
& RCV& $-0.495$ (0.451) & $-4.303$ (0.362) & $-7.245$ (0.286) \\
%& EBIC& -0.662 (0.333) & -0.412 (0.255) & -0.134 (0.203) \\
& oracle& $-0.316$ (0.328) & $-0.283$ (0.254) & $-0.021$ (0.201) \\\hline
        \multirow{3}{*}{\begin{minipage}{2cm} $b=1/\sqrt{3}$ \\$\rho=0.5$ \end{minipage}}
& \fidu & 0.352 (0.335) & 0.271 (0.285) & 1.046 (0.227) \\
& RCV& 0.455 (0.467) & $-10.333$ (0.334) & $-17.287$ (0.247) \\
%& EBIC& 0.242 (0.338) & 0.693 (0.3) & 1.317 (0.24) \\
& oracle& 0.367 (0.329) & $-0.548$ (0.258) & $-0.406$ (0.205) \\\hline
        \multirow{3}{*}{\begin{minipage}{2cm} $b=2/\sqrt{3}$ \\$\rho=0.5$ \end{minipage}}
& \fidu & $-0.505$ (0.328) & $-0.092$ (0.263) & $-0.302$ (0.199) \\
& RCV& $-0.533$ (0.442) & $-3.046$ (0.357) & $-6.73$ (0.257) \\
%& EBIC& -0.602 (0.33) & -0.295 (0.263) & -0.519 (0.199) \\
& oracle& $-0.103$ (0.325) & $-0.160$ (0.261) & $-0.483$ (0.198) \\\hline
        \multirow{3}{*}{\begin{minipage}{2cm} $b=3/\sqrt{3}$ \\$\rho=0.5$ \end{minipage}}
& \fidu & $-1.585$ (0.304) & 0.135 (0.259) & $-0.080$ (0.198) \\
& RCV& $-1.404$ (0.430) & $-2.275$ (0.342) & $-3.279$ (0.274) \\
%& EBIC& -1.69 (0.304) & -0.228 (0.259) & -0.383 (0.199) \\
& oracle& $-1.251$ (0.302) & $-0.188$ (0.258) & $-0.355$ (0.197) \\\hline
    \end{tabular}
    \caption{ Bias of the various estimates of $\sigma^2$.  Numbers in parentheses are standard errors, reported in \%.}
\label{table:sigmaestimate}
\end{table}

We also obtained two sets of 90\%, 95\% and 99\% confidence intervals for $\sigma^2$ from each simulated data set.  The first set was computed using the proposed generalized fiducial method, and the second was calculated by applying classical theory to the true model.  Again, the last method cannot be used in practice, and is used for benchmark comparisons; i.e., the oracle method.  The empirical coverage rates of these confidence intervals are summarized in Table~\ref{table:sci1}.  It can be seen that the generalized fiducial confidence intervals are nearly as good as the oracle confidence intervals.

\begin{table}[ht]
    \centering
    \renewcommand{\baselinestretch}{1.2}
    \scriptsize
    \begin{tabular}{|c|c|c|ccc|}\hline
    &    \multicolumn{2}{c|}{}  & 90\% & 95\% & 99\% \\\hline
    \multirow{12}{*}{$(n,p,d)=(200,2000,3)$} &   \multirow{2}{*}{\begin{minipage}{2cm} $b=1/\sqrt{5}$ \\$\rho=0$ \end{minipage}}& \fidu& 0.895 (0.338) & 0.949 (0.405) & 0.985 (0.537) \\
%&& RCV& 0.744 (0.334) & 0.818 (0.398) & 0.921 (0.523) \\
%&& EBIC& 0.887 (0.335) & 0.943 (0.4) & 0.984 (0.532) \\
&& oracle& 0.896 (0.336) & 0.948 (0.402) & 0.985 (0.534) \\\cline{2-6}
 &       \multirow{2}{*}{\begin{minipage}{2cm} $b=2/\sqrt{5}$ \\$\rho=0$ \end{minipage}}& \fidu& 0.892 (0.337) & 0.937 (0.404) & 0.987 (0.535) \\
%&& RCV& 0.724 (0.328) & 0.802 (0.391) & 0.908 (0.514) \\
%&& EBIC& 0.878 (0.334) & 0.933 (0.399) & 0.987 (0.53) \\
&& oracle& 0.892 (0.335) & 0.941 (0.401) & 0.988 (0.532) \\\cline{2-6}
 &       \multirow{2}{*}{\begin{minipage}{2cm} $b=3/\sqrt{5}$ \\$\rho=0$ \end{minipage}}& \fidu& 0.884 (0.338) & 0.941 (0.404) & 0.986 (0.536) \\
%&& RCV& 0.756 (0.327) & 0.814 (0.39) & 0.918 (0.513) \\
%&& EBIC& 0.881 (0.334) & 0.938 (0.4) & 0.983 (0.531) \\
&& oracle& 0.886 (0.335) & 0.943 (0.401) & 0.986 (0.533) \\\cline{2-6}
   &     \multirow{2}{*}{\begin{minipage}{2cm} $b=1/\sqrt{5}$ \\$\rho=0.5$ \end{minipage}}& \fidu& 0.895 (0.344) & 0.945 (0.412) & 0.988 (0.547) \\
%&& RCV& 0.747 (0.33) & 0.814 (0.394) & 0.911 (0.518) \\
%& EBIC& 0.889 (0.337) & 0.939 (0.403) & 0.986 (0.536) \\
&& oracle& 0.896 (0.338) & 0.946 (0.404) & 0.988 (0.536) \\\cline{2-6}
 &       \multirow{2}{*}{\begin{minipage}{2cm} $b=2/\sqrt{5}$ \\$\rho=0.5$ \end{minipage}}& \fidu& 0.889 (0.339) & 0.939 (0.405) & 0.991 (0.538) \\
%&& RCV& 0.748 (0.327) & 0.839 (0.39) & 0.926 (0.512) \\
%&& EBIC& 0.884 (0.334) & 0.935 (0.4) & 0.989 (0.531) \\
&& oracle& 0.891 (0.336) & 0.94 (0.402) & 0.991 (0.534) \\\cline{2-6}
 &       \multirow{2}{*}{\begin{minipage}{2cm} $b=3/\sqrt{5}$ \\$\rho=0.5$ \end{minipage}}& \fidu& 0.906 (0.335) & 0.955 (0.401) & 0.993 (0.532) \\
%&& RCV& 0.758 (0.324) & 0.832 (0.386) & 0.923 (0.508) \\
%&& EBIC& 0.905 (0.331) & 0.954 (0.396) & 0.992 (0.525) \\
&& oracle& 0.908 (0.332) & 0.957 (0.397) & 0.992 (0.528) \\\hline
  \multirow{12}{*}{$(n,p,d)=(300,8000,5)$} &       \multirow{2}{*}{\begin{minipage}{2cm} $b=1/\sqrt{5}$ \\$\rho=0$ \end{minipage}}& \fidu& 0.891 (0.277) & 0.948 (0.331) & 0.985 (0.438) \\
%&& RCV& 0.315 (0.224) & 0.38 (0.266) & 0.51 (0.35) \\
%&& EBIC& 0.877 (0.273) & 0.935 (0.326) & 0.984 (0.431) \\
&& oracle& 0.898 (0.273) & 0.948 (0.326) & 0.987 (0.432) \\\cline{2-6}
 &       \multirow{2}{*}{\begin{minipage}{2cm} $b=2/\sqrt{5}$ \\$\rho=0$ \end{minipage}}& \fidu& 0.909 (0.275) & 0.951 (0.328) & 0.987 (0.434) \\
%&& RCV& 0.594 (0.247) & 0.671 (0.295) & 0.781 (0.387) \\
%&& EBIC& 0.903 (0.272) & 0.95 (0.325) & 0.985 (0.43) \\
&& oracle& 0.904 (0.272) & 0.95 (0.325) & 0.985 (0.43) \\\cline{2-6}
&        \multirow{2}{*}{\begin{minipage}{2cm} $b=3/\sqrt{5}$ \\$\rho=0$ \end{minipage}}& \fidu& 0.913 (0.274) & 0.953 (0.328) & 0.993 (0.433) \\
%&& RCV& 0.674 (0.257) & 0.767 (0.306) & 0.877 (0.403) \\
%&& EBIC& 0.904 (0.272) & 0.953 (0.326) & 0.993 (0.431) \\
&& oracle& 0.907 (0.273) & 0.955 (0.326) & 0.993 (0.431) \\\cline{2-6}
&        \multirow{2}{*}{\begin{minipage}{2cm} $b=1/\sqrt{5}$ \\$\rho=0.5$ \end{minipage}}& \fidu& 0.887 (0.286) & 0.936 (0.342) & 0.984 (0.453) \\
%&& RCV& 0.525 (0.241) & 0.607 (0.287) & 0.75 (0.377) \\
%&& EBIC& 0.856 (0.275) & 0.912 (0.329) & 0.974 (0.435) \\
&& oracle& 0.898 (0.272) & 0.948 (0.325) & 0.992 (0.43) \\\cline{2-6}
 &       \multirow{2}{*}{\begin{minipage}{2cm} $b=2/\sqrt{5}$ \\$\rho=0.5$ \end{minipage}}& \fidu& 0.894 (0.275) & 0.947 (0.328) & 0.99 (0.434) \\
%&& RCV& 0.724 (0.26) & 0.791 (0.31) & 0.888 (0.408) \\
%&& EBIC& 0.89 (0.273) & 0.945 (0.326) & 0.99 (0.431) \\
&& oracle& 0.893 (0.273) & 0.946 (0.326) & 0.992 (0.432) \\\cline{2-6}
 &       \multirow{2}{*}{\begin{minipage}{2cm} $b=3/\sqrt{5}$ \\$\rho=0.5$ \end{minipage}}& \fidu& 0.906 (0.274) & 0.954 (0.328) & 0.99 (0.433) \\
%&& RCV& 0.761 (0.262) & 0.833 (0.313) & 0.918 (0.411) \\
%&& EBIC& 0.906 (0.273) & 0.951 (0.326) & 0.99 (0.432) \\
&& oracle& 0.906 (0.273) & 0.952 (0.326) & 0.99 (0.432) \\\hline
  \multirow{12}{*}{$(n,p,d)=(500,50000,8)$} &       \multirow{2}{*}{\begin{minipage}{2cm} $b=1/\sqrt{5}$ \\$\rho=0$ \end{minipage}}& \fidu& 0.88 (0.215) & 0.939 (0.257) & 0.989 (0.339) \\
%&& RCV& 0.006 (0.151) & 0.012 (0.18) & 0.035 (0.237) \\
%&& EBIC& 0.857 (0.211) & 0.922 (0.252) & 0.984 (0.333) \\
&& oracle& 0.909 (0.211) & 0.952 (0.252) & 0.99 (0.332) \\\cline{2-6}
  &      \multirow{2}{*}{\begin{minipage}{2cm} $b=2/\sqrt{5}$ \\$\rho=0$ \end{minipage}}& \fidu& 0.898 (0.212) & 0.942 (0.253) & 0.991 (0.333) \\
%&& RCV& 0.276 (0.179) & 0.344 (0.213) & 0.492 (0.281) \\
%&& EBIC& 0.896 (0.21) & 0.941 (0.251) & 0.991 (0.332) \\
&& oracle& 0.899 (0.211) & 0.942 (0.251) & 0.991 (0.332) \\\cline{2-6}
 &       \multirow{2}{*}{\begin{minipage}{2cm} $b=3/\sqrt{5}$ \\$\rho=0$ \end{minipage}}& \fidu& 0.901 (0.212) & 0.952 (0.253) & 0.991 (0.333) \\
%&& RCV& 0.579 (0.193) & 0.655 (0.23) & 0.774 (0.302) \\
%&& EBIC& 0.897 (0.211) & 0.949 (0.251) & 0.99 (0.332) \\
&& oracle& 0.9 (0.211) & 0.953 (0.252) & 0.992 (0.332) \\\cline{2-6}
  &      \multirow{2}{*}{\begin{minipage}{2cm} $b=1/\sqrt{5}$ \\$\rho=0.5$ \end{minipage}}& \fidu& 0.865 (0.224) & 0.935 (0.267) & 0.985 (0.352) \\
%&& RCV& 0.165 (0.172) & 0.195 (0.205) & 0.315 (0.269) \\
%&& EBIC& 0.82 (0.214) & 0.9 (0.255) & 0.975 (0.337) \\
&& oracle& 0.9 (0.21) & 0.94 (0.251) & 0.99 (0.331) \\\cline{2-6}
  &      \multirow{2}{*}{\begin{minipage}{2cm} $b=2/\sqrt{5}$ \\$\rho=0.5$ \end{minipage}}& \fidu& 0.895 (0.211) & 0.95 (0.252) & 0.993 (0.332) \\
%&& RCV& 0.612 (0.194) & 0.682 (0.231) & 0.801 (0.304) \\
%&& EBIC& 0.894 (0.21) & 0.948 (0.251) & 0.991 (0.331) \\
&& oracle& 0.895 (0.21) & 0.949 (0.251) & 0.992 (0.331) \\\cline{2-6}
  &      \multirow{2}{*}{\begin{minipage}{2cm} $b=3/\sqrt{5}$ \\$\rho=0.5$ \end{minipage}}& \fidu& 0.905 (0.211) & 0.947 (0.251) & 0.989 (0.331) \\
%&& RCV& 0.719 (0.201) & 0.772 (0.24) & 0.869 (0.315) \\
%&& EBIC& 0.902 (0.21) & 0.945 (0.251) & 0.988 (0.331) \\
&& oracle& 0.903 (0.21) & 0.945 (0.251) & 0.99 (0.331) \\\hline
    \end{tabular}
    \caption{Empirical coverage rates for various confidence intervals for $\sigma^2$.  Numbers in parentheses are averaged widths of the confidence intervals.}
\label{table:sci1}
\end{table}

Lastly, for each simulated data set we applied three methods to compute the confidence intervals for the regression coefficients $\beta_j$'s and the mean function $E(Y_i|\bx_i)$ evaluated at 50 randomly selected design points $\bx_i$'s.  The three methods are the proposed generalized fiducial method, the RCV method of \citet{Fan2011}, and the oracle method that uses the true model.  As before the empirical coverage rates of these confidence intervals are calculated and they are reported in Tables~\ref{table:beta} and~\ref{table:mean}.  Note that only the confidence intervals for $\beta_1$ are reported, as the confidence intervals for other $\beta_j$'s have similar coverage rates.  Overall one can see that the generalized fiducial method gave quite reliable results, except for a few experimental settings where the confidence intervals were over-liberal.

In an attempt to produce a single summary statistic for comparing the empirical coverage rates of the confidence intervals produced by different methods, the following calculation has been done.  For all the 90\% generalized fiducial confidence intervals for $\beta_1$, we counted the number of times that their empirical coverage rates are within the range $(1-\alpha) \pm 1.96 \sqrt{\alpha(1-\alpha)/N_{\rm sim}}$, where $\alpha=0.10$ and $N_{\rm sim}=1000$ is the number of repetitions performed for each experimental setting.  Similar calculations were then performed for the 95\% and 99\% (i.e., $\alpha=0.05$ and $\alpha=0.01$) confidence intervals.  And it turns out that, for the proposed generalized fiducial method, out of the 54 empirical coverage rates, 33 of them are within their corresponding target ranges.  We have also done the same calculations for the RCV and the oracle methods, and the numbers of their empirical coverage rates that are inside their target ranges are, respectively, 17 and 50.  Lastly, we repeated the same calculations for the empirical coverage rates for $E(Y_i|\bx_i)$, and the corresponding numbers for the proposed, RCV and oracle methods are, respectively, 44, 23 and 54.  Of course, these numbers are not perfect for judging the relative merits of the different methods, but they seem to suggest that the proposed generalized fiducial method provides improvement over the RCV method.

\begin{table}[ht]
    \centering
    \renewcommand{\baselinestretch}{1}
    \scriptsize
    \begin{tabular}{|c|c|c|ccc|}\hline
    &    \multicolumn{2}{c|}{}  & 90\% & 95\% & 99\% \\\hline
        \multirow{18}{*}{$(n,p,d)=(200,2000,3)$}&    \multirow{3}{*}{\begin{minipage}{2cm} $b=1/\sqrt{3}$ \\$\rho=0$ \end{minipage}}
& \fidu & 0.888 (0.236) & 0.946 (0.283) & 0.987 (0.377) \\
&& RCV& 0.869 (0.250) & 0.915 (0.298) & 0.956 (0.392) \\
%&& EBIC& 0.892 (0.233) & 0.944 (0.278) & 0.987 (0.365) \\
&& oracle& 0.897 (0.235) & 0.946 (0.279) & 0.988 (0.367) \\\cline{2-6}
 &       \multirow{3}{*}{\begin{minipage}{2cm} $b=2/\sqrt{3}$ \\$\rho=0$ \end{minipage}}
& \fidu & 0.884 (0.235) & 0.948 (0.282) & 0.991 (0.376) \\
&& RCV& 0.887 (0.238) & 0.945 (0.284) & 0.988 (0.373) \\
%&& EBIC& 0.887 (0.234) & 0.945 (0.278) & 0.989 (0.366) \\
&& oracle& 0.889 (0.234) & 0.946 (0.279) & 0.990 (0.367) \\\cline{2-6}
 &       \multirow{3}{*}{\begin{minipage}{2cm} $b=3/\sqrt{3}$ \\$\rho=0$ \end{minipage}}
& \fidu & 0.892 (0.236) & 0.947 (0.282) & 0.987 (0.376) \\
&& RCV& 0.896 (0.238) & 0.95 (0.284) & 0.99 (0.373) \\
%&& EBIC& 0.897 (0.234) & 0.951 (0.278) & 0.987 (0.366) \\
&& oracle& 0.897 (0.234) & 0.952 (0.279) & 0.987 (0.367) \\\cline{2-6}
 &       \multirow{3}{*}{\begin{minipage}{2cm} $b=1/\sqrt{3}$ \\$\rho=0.5$ \end{minipage}}
& \fidu & 0.886 (0.282) & 0.936 (0.338) & 0.985 (0.454) \\
&& RCV& 0.814 (0.289) & 0.849 (0.345) & 0.902 (0.453) \\
%&& EBIC& 0.875 (0.268) & 0.925 (0.32) & 0.969 (0.42) \\
&& oracle& 0.894 (0.271) & 0.943 (0.323) & 0.988 (0.424) \\\cline{2-6}
 &       \multirow{3}{*}{\begin{minipage}{2cm} $b=2/\sqrt{3}$ \\$\rho=0.5$ \end{minipage}}
& \fidu & 0.898 (0.271) & 0.944 (0.325) & 0.987 (0.433) \\
&& RCV& 0.903 (0.274) & 0.945 (0.326) & 0.988 (0.429) \\
%&& EBIC& 0.892 (0.27) & 0.948 (0.321) & 0.986 (0.422) \\
&& oracle& 0.894 (0.270) & 0.949 (0.322) & 0.986 (0.423) \\\cline{2-6}
 &       \multirow{3}{*}{\begin{minipage}{2cm} $b=3/\sqrt{3}$ \\$\rho=0.5$ \end{minipage}}
& \fidu & 0.901 (0.269) & 0.948 (0.322) & 0.989 (0.429) \\
&& RCV& 0.899 (0.271) & 0.953 (0.323) & 0.988 (0.424) \\
%&& EBIC& 0.897 (0.269) & 0.952 (0.32) & 0.99 (0.421) \\
&& oracle& 0.897 (0.269) & 0.955 (0.321) & 0.99 (0.422) \\\hline
        \multirow{18}{*}{$(n,p,d)=(300,8000,5)$}&        \multirow{3}{*}{\begin{minipage}{2cm} $b=1/\sqrt{5}$ \\$\rho=0$ \end{minipage}}
& \fidu & 0.810 (0.191) & 0.896 (0.229) & 0.976 (0.303) \\
&& RCV& 0.903 (0.204) & 0.935 (0.243) & 0.956 (0.320) \\
%&& EBIC& 0.893 (0.191) & 0.942 (0.228) & 0.987 (0.299) \\
&& oracle& 0.900 (0.192) & 0.948 (0.229) & 0.992 (0.301) \\\cline{2-6}
 &       \multirow{3}{*}{\begin{minipage}{2cm} $b=2/\sqrt{5}$ \\$\rho=0$ \end{minipage}}
& \fidu & 0.871 (0.189) & 0.936 (0.226) & 0.984 (0.300) \\
&& RCV& 0.897 (0.201) & 0.936 (0.239) & 0.981 (0.315) \\
%&& EBIC& 0.907 (0.191) & 0.959 (0.228) & 0.989 (0.3) \\
&& oracle& 0.907 (0.191) & 0.959 (0.228) & 0.989 (0.300) \\\cline{2-6}
 &       \multirow{3}{*}{\begin{minipage}{2cm} $b=3/\sqrt{5}$ \\$\rho=0$ \end{minipage}}
& \fidu & 0.888 (0.19) & 0.934 (0.227) & 0.984 (0.301) \\
&& RCV& 0.900 (0.197) & 0.945 (0.235) & 0.979 (0.309) \\
%&& EBIC& 0.879 (0.192) & 0.941 (0.228) & 0.991 (0.3) \\
&& oracle& 0.879 (0.192) & 0.941 (0.228) & 0.991 (0.300) \\\cline{2-6}
 &       \multirow{3}{*}{\begin{minipage}{2cm} $b=1/\sqrt{5}$ \\$\rho=0.5$ \end{minipage}}
& \fidu & 0.812 (0.269) & 0.887 (0.322) & 0.963 (0.427) \\
&& RCV& 0.871 (0.236) & 0.915 (0.281) & 0.960 (0.369) \\
%&& EBIC& 0.853 (0.213) & 0.89 (0.254) & 0.933 (0.334) \\
&& oracle& 0.912 (0.221) & 0.954 (0.264) & 0.992 (0.346) \\\cline{2-6}
 &       \multirow{3}{*}{\begin{minipage}{2cm} $b=2/\sqrt{5}$ \\$\rho=0.5$ \end{minipage}}
& \fidu & 0.895 (0.250) & 0.949 (0.299) & 0.989 (0.396) \\
&& RCV& 0.864 (0.224) & 0.922 (0.266) & 0.975 (0.350) \\
%&& EBIC& 0.891 (0.222) & 0.949 (0.264) & 0.991 (0.347) \\
&& oracle& 0.891 (0.222) & 0.950 (0.264) & 0.991 (0.347) \\\cline{2-6}
 &       \multirow{3}{*}{\begin{minipage}{2cm} $b=3/\sqrt{5}$ \\$\rho=0.5$ \end{minipage}}
& \fidu & 0.908 (0.250) & 0.950 (0.299) & 0.990 (0.397) \\
&& RCV& 0.852 (0.220) & 0.917 (0.262) & 0.975 (0.344) \\
%&& EBIC& 0.903 (0.222) & 0.949 (0.264) & 0.983 (0.347) \\
&& oracle& 0.904 (0.222) & 0.949 (0.264) & 0.983 (0.347) \\\hline
        \multirow{18}{*}{$(n,p,d)=(500,50000,8)$}&       \multirow{3}{*}{\begin{minipage}{2cm} $b=1/\sqrt{8}$ \\$\rho=0$ \end{minipage}}
& \fidu & 0.781 (0.148) & 0.875 (0.177) & 0.978 (0.233) \\
&& RCV& 0.813 (0.151) & 0.857 (0.180) & 0.884 (0.237) \\
%&& EBIC& 0.902 (0.147) & 0.94 (0.175) & 0.985 (0.23) \\
&& oracle& 0.910 (0.149) & 0.954 (0.177) & 0.993 (0.233) \\\cline{2-6}
 &       \multirow{3}{*}{\begin{minipage}{2cm} $b=2/\sqrt{8}$ \\$\rho=0$ \end{minipage}}
& \fidu & 0.853 (0.147) & 0.919 (0.176) & 0.980 (0.232) \\
&& RCV& 0.804 (0.156) & 0.878 (0.186) & 0.965 (0.244) \\
%&& EBIC& 0.901 (0.148) & 0.947 (0.177) & 0.988 (0.232) \\
&& oracle& 0.902 (0.148) & 0.947 (0.177) & 0.988 (0.232) \\\cline{2-6}
 &       \multirow{3}{*}{\begin{minipage}{2cm} $b=3/\sqrt{8}$ \\$\rho=0$ \end{minipage}}
& \fidu & 0.873 (0.147) & 0.925 (0.176) & 0.986 (0.232) \\
&& RCV& 0.841 (0.155) & 0.911 (0.184) & 0.981 (0.242) \\
%&& EBIC& 0.897 (0.148) & 0.944 (0.177) & 0.988 (0.232) \\
&& oracle& 0.897 (0.149) & 0.944 (0.177) & 0.988 (0.233) \\\cline{2-6}
 &       \multirow{3}{*}{\begin{minipage}{2cm} $b=1/\sqrt{8}$ \\$\rho=0.5$ \end{minipage}}
& \fidu & 0.820 (0.206) & 0.885 (0.246) & 0.950 (0.324) \\
&& RCV& 0.895 (0.179) & 0.935 (0.213) & 0.965 (0.280) \\
%&& EBIC& 0.815 (0.166) & 0.865 (0.197) & 0.885 (0.259) \\
&& oracle& 0.925 (0.172) & 0.965 (0.204) & 0.995 (0.269) \\\cline{2-6}
 &       \multirow{3}{*}{\begin{minipage}{2cm} $b=2/\sqrt{8}$ \\$\rho=0.5$ \end{minipage}}
& \fidu & 0.897 (0.193) & 0.949 (0.230) & 0.988 (0.304) \\
&& RCV& 0.861 (0.169) & 0.922 (0.202) & 0.976 (0.265) \\
%&& EBIC& 0.893 (0.171) & 0.942 (0.204) & 0.989 (0.268) \\
&& oracle& 0.893 (0.171) & 0.944 (0.204) & 0.989 (0.268) \\\cline{2-6}
 &       \multirow{3}{*}{\begin{minipage}{2cm} $b=3/\sqrt{8}$ \\$\rho=0.5$ \end{minipage}}
& \fidu & 0.888 (0.193) & 0.945 (0.230) & 0.989 (0.304) \\
&& RCV& 0.840 (0.168) & 0.909 (0.201) & 0.968 (0.264) \\
%&& EBIC& 0.899 (0.171) & 0.942 (0.204) & 0.987 (0.268) \\
&& oracle& 0.899 (0.171) & 0.942 (0.204) & 0.987 (0.268) \\\hline
    \end{tabular}
    \caption{Empirical coverage rates for the confidence intervals for $\beta_1$.  The numbers in the parentheses are the averaged widths of the corresponding confidence intervals.}
\label{table:beta}
\end{table}

\begin{table}[ht]
    \centering
    \renewcommand{\baselinestretch}{1}
    \scriptsize
    \begin{tabular}{|c|c|c|ccc|}\hline
      &  \multicolumn{2}{c|}{}  & 90\% & 95\% & 99\% \\\hline
             \multirow{18}{*}{$(n,p,d)=(200,2000,3)$}&    \multirow{3}{*}{\begin{minipage}{2cm} $b=1/\sqrt{3}$ \\$\rho=0$ \end{minipage}}
& \fidu & 0.899 (0.421) & 0.948 (0.511) & 0.988 (0.696) \\
&& RCV& 0.966 (1.160) & 0.981 (1.382) & 0.993 (1.817) \\
%&& EBIC& 0.865 (0.346) & 0.918 (0.412) & 0.967 (0.542) \\
&& oracle& 0.896 (0.343) & 0.947 (0.409) & 0.989 (0.538) \\\cline{2-6}
&        \multirow{3}{*}{\begin{minipage}{2cm} $b=2/\sqrt{3}$ \\$\rho=0$ \end{minipage}}
& \fidu & 0.903 (0.424) & 0.953 (0.516) & 0.990 (0.704) \\
&& RCV& 0.857 (0.603) & 0.910 (0.718) & 0.966 (0.944) \\
%&& EBIC& 0.868 (0.344) & 0.924 (0.41) & 0.972 (0.539) \\
&& oracle& 0.888 (0.342) & 0.944 (0.408) & 0.988 (0.536) \\\cline{2-6}
 &       \multirow{3}{*}{\begin{minipage}{2cm} $b=3/\sqrt{3}$ \\$\rho=0$ \end{minipage}}
& \fidu & 0.911 (0.428) & 0.956 (0.519) & 0.991 (0.709) \\
&& RCV& 0.931 (0.605) & 0.965 (0.720) & 0.992 (0.947) \\
%&& EBIC& 0.878 (0.345) & 0.929 (0.411) & 0.973 (0.54) \\
&& oracle& 0.897 (0.343) & 0.947 (0.409) & 0.987 (0.537) \\\cline{2-6}
 &       \multirow{3}{*}{\begin{minipage}{2cm} $b=1/\sqrt{3}$ \\$\rho=0.5$ \end{minipage}}
& \fidu & 0.903 (0.452) & 0.948 (0.549) & 0.987 (0.748) \\
&& RCV& 0.925 (1.281) & 0.943 (1.526) & 0.964 (2.005) \\
%&& EBIC& 0.853 (0.343) & 0.902 (0.409) & 0.947 (0.537) \\
&& oracle& 0.892 (0.344) & 0.944 (0.410) & 0.987 (0.538) \\\cline{2-6}
  &      \multirow{3}{*}{\begin{minipage}{2cm} $b=2/\sqrt{3}$ \\$\rho=0.5$ \end{minipage}}
& \fidu & 0.910 (0.444) & 0.955 (0.538) & 0.990 (0.733) \\
&& RCV& 0.855 (0.583) & 0.907 (0.695) & 0.963 (0.914) \\
%&& EBIC& 0.87 (0.345) & 0.923 (0.411) & 0.967 (0.541) \\
&& oracle& 0.896 (0.343) & 0.948 (0.408) & 0.988 (0.536) \\\cline{2-6}
  &      \multirow{3}{*}{\begin{minipage}{2cm} $b=3/\sqrt{3}$ \\$\rho=0.5$ \end{minipage}}
& \fidu & 0.913 (0.438) & 0.959 (0.532) & 0.993 (0.725) \\
&& RCV& 0.925 (0.492) & 0.961 (0.587) & 0.993 (0.771) \\
%&& EBIC& 0.874 (0.344) & 0.925 (0.41) & 0.972 (0.539) \\
&& oracle& 0.899 (0.342) & 0.947 (0.408) & 0.989 (0.536) \\\hline
        \multirow{18}{*}{$(n,p,d)=(300,8000,5)$}&        \multirow{3}{*}{\begin{minipage}{2cm} $b=1/\sqrt{5}$ \\$\rho=0$ \end{minipage}}
& \fidu & 0.888 (0.444) & 0.938 (0.536) & 0.981 (0.725) \\
&& RCV& 0.951 (1.864) & 0.973 (2.221) & 0.99 (2.919) \\
%&& EBIC& 0.865 (0.389) & 0.916 (0.464) & 0.961 (0.609) \\
&& oracle& 0.898 (0.388) & 0.950 (0.462) & 0.990 (0.607) \\\cline{2-6}
 &       \multirow{3}{*}{\begin{minipage}{2cm} $b=2/\sqrt{5}$ \\$\rho=0$ \end{minipage}}
& \fidu & 0.909 (0.439) & 0.956 (0.531) & 0.992 (0.724) \\
&& RCV& 0.949 (1.291) & 0.977 (1.538) & 0.995 (2.022) \\
%&& EBIC& 0.89 (0.387) & 0.94 (0.461) & 0.984 (0.606) \\
&& oracle& 0.900 (0.386) & 0.949 (0.46) & 0.990 (0.605) \\\cline{2-6}
 &       \multirow{3}{*}{\begin{minipage}{2cm} $b=3/\sqrt{5}$ \\$\rho=0$ \end{minipage}}
& \fidu & 0.909 (0.429) & 0.957 (0.519) & 0.992 (0.708) \\
&& RCV& 0.942 (0.915) & 0.973 (1.090) & 0.995 (1.432) \\
%&& EBIC& 0.89 (0.388) & 0.941 (0.462) & 0.984 (0.607) \\
&& oracle& 0.897 (0.387) & 0.948 (0.461) & 0.990 (0.606) \\\cline{2-6}
 &       \multirow{3}{*}{\begin{minipage}{2cm} $b=1/\sqrt{5}$ \\$\rho=0.5$ \end{minipage}}
& \fidu & 0.871 (0.496) & 0.925 (0.602) & 0.975 (0.820) \\
&& RCV& 0.953 (1.641) & 0.978 (1.956) & 0.996 (2.570) \\
%&& EBIC& 0.78 (0.378) & 0.825 (0.45) & 0.877 (0.591) \\
&& oracle& 0.898 (0.387) & 0.947 (0.461) & 0.988 (0.606) \\\cline{2-6}
 &       \multirow{3}{*}{\begin{minipage}{2cm} $b=2/\sqrt{5}$ \\$\rho=0.5$ \end{minipage}}
& \fidu & 0.914 (0.437) & 0.962 (0.531) & 0.994 (0.728) \\
&& RCV& 0.947 (0.741) & 0.977 (0.883) & 0.996 (1.160) \\
%&& EBIC& 0.892 (0.387) & 0.946 (0.462) & 0.986 (0.607) \\
&& oracle& 0.901 (0.387) & 0.954 (0.461) & 0.991 (0.606) \\\cline{2-6}
&        \multirow{3}{*}{\begin{minipage}{2cm} $b=3/\sqrt{5}$ \\$\rho=0.5$ \end{minipage}}
& \fidu & 0.914 (0.422) & 0.960 (0.512) & 0.993 (0.701) \\
&& RCV& 0.914 (0.431) & 0.958 (0.514) & 0.992 (0.676) \\
%&& EBIC& 0.897 (0.388) & 0.949 (0.462) & 0.989 (0.607) \\
&& oracle& 0.900 (0.388) & 0.951 (0.462) & 0.991 (0.607) \\\hline
        \multirow{18}{*}{$(n,p,d)=(500,50000,8)$}&   \multirow{3}{*}{\begin{minipage}{2cm} $b=1/\sqrt{8}$ \\$\rho=0$ \end{minipage}}
& \fidu & 0.841 (0.445) & 0.896 (0.534) & 0.951 (0.711) \\
&& RCV& 0.934 (1.889) & 0.960 (2.251) & 0.983 (2.958) \\
%&& EBIC& 0.813 (0.41) & 0.867 (0.489) & 0.924 (0.642) \\
&& oracle& 0.902 (0.409) & 0.953 (0.488) & 0.991 (0.641) \\\cline{2-6}
 &       \multirow{3}{*}{\begin{minipage}{2cm} $b=2/\sqrt{8}$ \\$\rho=0$ \end{minipage}}
& \fidu & 0.907 (0.435) & 0.955 (0.522) & 0.991 (0.697) \\
&& RCV& 0.951 (1.573) & 0.980 (1.874) & 0.997 (2.463) \\
%&& EBIC& 0.897 (0.409) & 0.946 (0.487) & 0.986 (0.641) \\
&& oracle& 0.903 (0.409) & 0.951 (0.487) & 0.990 (0.640) \\\cline{2-6}
 &       \multirow{3}{*}{\begin{minipage}{2cm} $b=3/\sqrt{8}$ \\$\rho=0$ \end{minipage}}
& \fidu & 0.900 (0.429) & 0.951 (0.515) & 0.990 (0.687) \\
&& RCV& 0.957 (1.187) & 0.983 (1.415) & 0.998 (1.860) \\
%&& EBIC& 0.89 (0.41) & 0.942 (0.488) & 0.985 (0.641) \\
&& oracle& 0.898 (0.409) & 0.949 (0.488) & 0.989 (0.641) \\\cline{2-6}
 &       \multirow{3}{*}{\begin{minipage}{2cm} $b=1/\sqrt{8}$ \\$\rho=0.5$ \end{minipage}}
& \fidu & 0.829 (0.501) & 0.892 (0.601) & 0.958 (0.803) \\
&& RCV& 0.945 (1.713) & 0.978 (2.041) & 0.996 (2.682) \\
%&& EBIC& 0.709 (0.397) & 0.76 (0.473) & 0.834 (0.622) \\
&& oracle& 0.905 (0.408) & 0.951 (0.486) & 0.992 (0.639) \\\cline{2-6}
  &      \multirow{3}{*}{\begin{minipage}{2cm} $b=2/\sqrt{8}$ \\$\rho=0.5$ \end{minipage}}
& \fidu & 0.907 (0.430) & 0.956 (0.517) & 0.993 (0.693) \\
&& RCV& 0.951 (0.708) & 0.979 (0.844) & 0.997 (1.109) \\
%&& EBIC& 0.898 (0.409) & 0.948 (0.487) & 0.99 (0.64) \\
&& oracle& 0.900 (0.408) & 0.951 (0.487) & 0.992 (0.640) \\\cline{2-6}
  &      \multirow{3}{*}{\begin{minipage}{2cm} $b=3/\sqrt{8}$ \\$\rho=0.5$ \end{minipage}}
& \fidu & 0.903 (0.421) & 0.953 (0.505) & 0.991 (0.675) \\
&& RCV& 0.900 (0.417) & 0.949 (0.497) & 0.990 (0.653) \\
%&& EBIC& 0.896 (0.409) & 0.947 (0.487) & 0.988 (0.64) \\
&& oracle& 0.898 (0.409) & 0.949 (0.487) & 0.990 (0.640) \\\hline
    \end{tabular}
    \caption{Empirical coverage rates for the confidence intervals for $E(Y_i|\bx_i)$.  The numbers in the parentheses are the averaged widths of the corresponding confidence intervals.}
\label{table:mean}
\end{table}

\subsection{Real Data Example: Housing Price Appreciation}
This section analyses a data set that contains 119 months of housing price appreciation (HPA) of the national house price index (HPI) for 381 core-based statistical areas (CBSAs) in the united states.  Here HPA is defined as the percentage of monthly change in log-HPI for each of the 381 CBSAs.  The goal of the analysis is to predict future HPA values for these CBSAs using existing data.  This data set was recorded from 1996 to 2005, and has been studied for example by \citet{Fan2011}.

Of course, house prices depend on geographical locations and various macroeconomic factors.  As argued by \citet{Fan2011}, effects from macroeconomic factors can be well summarized by the national HPA.  Let $X_{t,j}$ be the HPA of the $j$-th CBSA in month $t$, and $X_{t,{\rm N}}$ be the national HPA of month $t$.  Then for any $k=1, \ldots, 381$, a reasonable model for a 1-year ahead HPA prediction for the $k$-th CBSA is
\[
X_{t,k} = \sum_{j=1}^{381} \beta^{(k)}_{j} X_{t-1,j} + \beta^{(k)}_{{\rm N}} X_{t-1,{\rm N}} + \epsilon_{t-1},
\]
where $\beta^{(k)}_{j}$'s and $\beta^{(k)}_{\rm N}$ are model parameters and $\epsilon_{t-1}$ is an independent random error.  Given the national HPA $X_{t-1,{\rm N}}$, it is reasonable to assume that areas that are far away would have minimal influence on the local house prices, therefore one can assume the $\beta^{(k)}_{j}$'s are sparse.  Note that for any given $k$, we have ``$p>n$'', as $p=382$ and $n=119$.

For illustrative purposes, we apply the proposed generalized fiducial procedure to the above model for one of the CBSAs: San Francisco-San Mateo-Redwood.  Two fitted models with non-negligible fiducial probabilities are returned: with probability 0.335 the housing appreciation of this area depends on itself and its nearby CBSA San Jose-San Francisco-Oakland, while with probability about 0.663, it depends only on the CBSA San Jose-San Francisco-Oakland.

We also obtained estimate for the noise standard deviation $\sigma$, which can be interpreted as a measure of prediction accuracy when forecasting the housing appreciation.  Our point estimate for $\sigma$ is 0.56 with a 95\% confidence as $(0.48, 0.65)$.  Our point estimate agrees with those reported in \citet{Fan2011}, although no confidence intervals are reported there.

\ignore{
\begin{table}[ht]
    \centering
    \footnotesize
    \begin{tabular}{|c|c|c|c|}\hline
CBSA & Model (probability) & Estimate of $\sigma$ & 95\% CI for $\sigma$\\ \hline
San Francisco-San Mateo-Redwood & \begin{minipage}{6cm} \vspace*{5pt} San Francisco-San Mateo-Redwood, \\ San Jose-San Francisco-Oakland (0.335) \vspace*{5pt} \end{minipage}& 0.56\% & $(0.48\%, 0.65\%)$\\\cline{2-2}
    &\begin{minipage}{6cm} \vspace*{5pt} San Jose-San Francisco-Oakland (0.663)    \vspace*{5pt}  \end{minipage}& &\\ \hline
Los Angeles-Long Beach -Glendale & \begin{minipage}{6cm} \vspace*{5pt} Los Angeles-Long Beach -Glendale \\Oxnard- Thousand Oaks-Ventura, \\ Riverside-San Bernardino-Ontario (0.730)\vspace*{5pt} \end{minipage}& 0.48\% & $(0.41\%, 0.57\%)$\\\cline{2-2}
    & \begin{minipage}{6cm} \vspace*{5pt} Los Angeles-Long Beach -Glendale \\ Riverside-San Bernardino-Ontario (0.184)   \vspace*{5pt}  \end{minipage}& &\\ \hline
Merced County, CA& \begin{minipage}{6cm} \vspace*{5pt} Merced County, CA (0.243)\vspace*{5pt} \end{minipage}& 0.65\% & $(0.56\%, 0.76\%)$\\\cline{2-2}
    & \begin{minipage}{6cm} \vspace*{5pt} Merced County, CA \\San Joaquin County, CA (0.483)   \vspace*{5pt}  \end{minipage}& &\\\cline{2-2}
    & \begin{minipage}{6cm} \vspace*{5pt} Merced County, CA \\Stanislaus County, CA (0.194)   \vspace*{5pt}  \end{minipage}& &\\ \hline

    \end{tabular}
    \caption{}
    \label{tab:}
\end{table}
}

\section{Conclusion}
\label{sec:conclude}
In this paper we studied the issue of uncertainty quantification in the ultrahigh dimensional regression problem.  We applied the generalized fiducial inference methodology to develop an inferential procedure for this problem.  Our theoretical results show that estimates obtained by this procedure are consistent, while confidence intervals constructed by this procedure are asymptotically correct in the frequentist sense.  Numerical results from simulation experiments confirm with these theoretical findings.  To the best of our knowledge, there are very few published papers that are devoted to quantify uncertainties in the ultrahigh dimensional regression problem, and hence the current paper is one of the first to provide a systematic treatment to this problem.  It also opens the possibility for using fiducial and related methods for conducting statistical inference for other ``large $p$ small $n$'' problems, such as classification and covariance matrix estimation.

%this is the first time that Fisher's fiducial idea is being applied to a high dimensional problem.  Thus this paper opens 

\appendix
\section{Derivation of~(\protect\ref{eq:FidModelProb})}
\label{app:derivegfd}
This appendix derives the generalized fiducial density~(\ref{eq:FidModelProb}).  A major challenge is to obtain a computable expression for the Jacobian~(\ref{eq:RecommendedJacobian1}).

First observe that the term $J(\by,\btheta)$ in~(\ref{eq:RecommendedJacobian1}) can be further simplified.  The product of Jacobian matrices in each of the summands of~(\ref{eq:RecommendedJacobian1}) simplifies to a matrix containing the $d$-columns of the $n\times d$ matrix $\left\{\frac{\bd}{\bd\by} \bG^{-1}(\by,\btheta)\right\}^{-1}\frac{\bd}{\bd\btheta} \bG^{-1}(\by,\btheta)$ and the $n-d$ columns of the identity matrix with columns $i_1,\ldots,i_d$ removed. Thus we have
\begin{equation}\label{eq:RecommendedJacobian2}
 J(\by,\btheta)=  \sum_{\substack{\bi=(i_1,\ldots,i_d) \\ 1\leq i_1<\cdots<i_d\leq n}}\left|\det\left[\left\{\frac{\bd}{\bd\by} \bG^{-1}(\by,\btheta)\right\}^{-1}\frac{\bd}{\bd\btheta} \bG^{-1}(\by,\btheta)\right]_\bi\right|,
\end{equation}
where for any $n\times d$ matrix $\bA$, the sub-matrix $(\bA)_\bi$ is the $d\times d$ matrix containing the rows $i_1,\ldots,i_d$ of $\bA$.

Then notice that each of the candidate model is a multiple regression model, with an implicit structural equation
\[
 \bY=\bG_M(\bbeta_M,\sigma^2,\bZ)=\bX_M \bbeta_M + \sigma \bZ,
\]
where $\bY$ is the observations, $\bX_M$ is the design matrix for model $M$, $\bbeta_M\in\mathbb R^{|M|}$ and $\sigma>0$ are parameters, and $\bZ$ is a vector of \iid standard normal random variables.  Plugging this into~(\ref{eq:RecommendedJacobian2}) and after some calculations one has
\[
J_M(\by, \btheta)=\sigma^{-2} \sum_{\substack{\bi=(i_0,\ldots,i_{|M|}) \\ 1\leq i_0<\cdots<i_{|M|}\leq n}}\left|\det\left(\by,\bX_M\right)_\bi\right|.
\]
Substituting this into~(\ref{eq:penalizedfiducia}) we have
\begin{equation}\label{eq:FidModelProbB}
    r(M)\propto \sum_{\substack{\bi=(i_0,\ldots,i_{|M|}) \\ 1\leq i_0<\cdots<i_{|M|}\leq n}}\left|\det\left(\by,\bX_M\right)_\bi\right|
 \Gamma\left( \frac{n-m}{2} \right) \left(\pi \text{RSS}_{M}\right)^{-\frac{n-m}{2}} |\det(\bX_{M}^T \bX_{M})|^{-\frac{1}{2}} e^{-q(M)},
\end{equation}
where $\text{RSS}_M $ denotes the residual sum of squares of model $M$ when the parameters are estimated using maximum likelihood, and the term $q(M)$ that controls the model dimension is given by~(\ref{eq:MDLpenalty}).

The expression~(\ref{eq:FidModelProbB}) has done well in our simulations.  However, the need for computing a sum of $\binom{n}{|M|+1}$ terms makes it very computationally expensive.  To seek for a faster alternative, we re-express the response $\bY$ for each fixed model as a column vector
\[\bv_M=[(\bX_M^T\bX_M)^{-1/2}\bX_M^T\by; (\text{RSS}_M)^{1/2}; \{\bI-\bX_M(\bX_M^T\bX_M)^{-1}\bX_M^T\}\by/\text{RSS}_M].\]
With this the Jacobian~(\ref{eq:RecommendedJacobian2}) becomes
\begin{align*}
 J_M^v(\by, \btheta)&=
 \sum_{\bi}\left|\det\left[\frac{\bd \bv_M}{\bd\by}  \left\{\frac{\bd}{\bd\by} \bG_M^{-1}(\by,\bbeta_M,\sigma^2)\right\}^{-1}\frac{\bd}{\bd(\by,\bbeta_M,\sigma^2)} \bG_M^{-1}(\by,\bbeta_M,\sigma^2)\right]_\bi\right|\\
 &=\sigma^{-2} |\det(\bX_M' \bX_M)|^{\frac{1}{2}} \text{RSS}_M^\frac{1}{2}.
\end{align*}
The simplification in the previous formula happens because all but the first $m+1$ rows of the matrix obtained as the product of matrices in the above expression are 0 and we therefore have only one non-zero determinant in the sum.  This together with the penalty~(\ref{eq:MDLpenalty}) brings us to the final generalized fiducial distribution~(\ref{eq:FidModelProb}).

Notice that both  $J_M^v(\by, \btheta)$ and $ J_M(\by, \btheta)$ are of the form
$C_M(\by) \sigma^{-2}$ where $C_M(\by)$ is a specific constant depending only on the observed data. Therefore the Jacobians can be viewed as improper Bayesian priors
$
 \pi(\bbeta_M,\sigma^{2}) \propto \sigma^{-2}.
$
As discussed in \cite{BergerPericchi2001} one the issues with the use of improper priors in Bayesian model selection is that a selection of a constant $C_B$ in the prior $C_B \sigma^{-2}$ is arbitrary. This is not a problem when a posterior with respect to one model is considered because the arbitrary constant cancels. However, it becomes a problem in model selection as the arbitrary constants $C_B$ influence the result making the use of improper prior for model selection difficult. Thus a contribution of fiducial inference is the choice of a particular constant $C_B$ for each of the model.
%Finally, we remark that if $n=m+1$ then
%$J_M^v(\by, \beta_M,\eta)=J_M(\by, \beta_M,\eta)$ and
%\[
% \int_{\mathbb R^{m}\times\mathbb R^+} f(\by,\beta_M,\eta) J(\bx,,\beta_M,\eta) d\eta d\beta_M=1,
%\]
%which is a property used by  \cite{BergerPericchi2001} to define their ``intrinsic Bayes factors''.
%

\section{Proof of Theorem~\protect\ref{th:main}}
\label{app:proofs}

\subsection{Lemmas}
First we present three lemmas, where detailed proofs can be found in \citet{luo2011}.  Lemma~\ref{lemma:1} is proved by applying Stirling's formula. Lemma~\ref{lemma:2} is proved by integration by parts and Lemma~\ref{lemma:3} is proved by applying Lemma~\ref{lemma:2}.

\begin{lemma} \label{lemma:1}
If $\log j / \log p \to \delta$ as $p \to \infty$, then
\[
\log\binom{p}{j} = j \log p (1-\delta) (1+o(1)).
\]
\end{lemma}

\begin{lemma} \label{lemma:2}
Let $\chi^2_j$ be a chi-square random variable with degrees of freedom $j$.  If $c \to \infty$ and $J/c \to 0$, then
\[
P(\chi^2_j > c) = \frac{1}{\Gamma(j/2)} (c/2)^{k/2-1} e^{-c/2} (1+o(1)),
\]
uniformly over $j \le J$.
\end{lemma}

\begin{lemma} \label{lemma:3}
Let $\chi^2_j$ be a chi-square random variable with degrees of freedom $j$. Let $c_j = 2 j \left\{ \log p + \log (j \log p) \right\}$. If $p \to \infty$, then for any $J \le p$,
\[
\sum_{j=1}^J \binom{p}{j} P(\chi^2_j > c_j ) \to 0.
\]
\end{lemma}

\subsection{Proof of Theorem~\protect\ref{th:main}}
This appendix presents the proof of Theorem~\ref{th:main}.  Some of the arguments are similar to those in \cite{luo2011}.

Denote $\mathcal M$ as the collection of models for which \eq{idCon} holds, i.e., $\mathcal M = \left\{ M : \ |M| \le k |M_0| \right\}$ for some fixed $k$.  We first prove that $\max_{\mathcal M} r(M)/r(M_0) \overset{P}{\to} 0$. WLOG, assume that $\sigma^2=1$. Let $m=|M|$ and $m_0=|M_0|$ whenever there is no ambiguity. Notice that $m_0=o(n)$ and $m=o(n)$. Rewrite
\[
r(M)/r(M_0) = \exp\left\{ -T_1 -T_2 \right\}
\]
where
\[
T_1 = \frac{n-m-1}{2} \log\left( \frac{\text{RSS}_{M}}{\text{RSS}_{M_0}}\right) ,
\]
\begin{align*}
    T_2 &= \frac{m-m_0}{2} \log n + \frac{m-m_0}{2} \log(\pi \text{RSS}_{M_0})+\log  \left\{ \Gamma\left( \frac{n-m_0}{2} \right) / \Gamma\left( \frac{n-m}{2} \right) \right\}\\
    & \quad - \gamma \log \binom{p}{m_0} + \gamma \log \binom{p}{m}.
\end{align*}

We are going to show that the followings hold uniformly for all $M$:
\begin{align}
&\begin{cases}
    T_1  =  \frac{\Delta_M(1+ o_p(1))}{2} & \quad \text{if } M_0 \not \subset M,
\\
T_2  \ge - \frac{3}{2} m_0 \log n- \gamma m_0 \log p& \quad \text{if } M_0 \not \subset M,
\end{cases}
\label{eq:a1}\\
&\begin{cases}
    T_1 \ge - (m-m_0) (1+\delta) \log p (1+o_p(1))&\quad \text{if } M_0 \subset M,\\
T_2 = \frac{3}{2}(m-m_0) \log n (1+o_p(1))+ \gamma (1-\delta) (m-m_0) \log p (1 + o(1))
& \quad \text{if } M_0 \subset M.
\end{cases}
\label{eq:a2}
\end{align}

\noindent {\bf Case 1: $M_0 \not\subset M$.}

Let $\mathcal M_j = \{M: \ |M| =j, M \in \mathcal M\}$.
First note that $\text{RSS}_{M_0} = (n-m_0) \left(1+o_p(1)  \right) = n \left( 1+o_p(1) \right)$,
\begin{align}
    \text{RSS}_{M}-\text{RSS}_{M_0} = \Delta(M)  + 2 \bmu^T \left( \bI - \bH_M \right) \bepsilon + \bepsilon^T \bH_M \bepsilon - \bepsilon^T \bH_{M_0} \bepsilon. \label{eq:rssdiff}
\end{align}
and $ \bepsilon^T \bH_{M_0} \bepsilon = m_0 (1+o_p(1))$.

Consider the second term in \eq{rssdiff} and denote $Z_M = \bmu^T \left( \bI - \bH_M \right) \bepsilon / \sqrt{\Delta_M}$, we have
\[
\bmu^T \left( \bI - \bH_M \right) \bepsilon = \sqrt{\Delta_M} Z_M
\]
and $Z_M \sim N(0,1)$. Let $c_j = 2 j \left\{ \log p + \log (j \log p) \right\}$. For simplicity, denote $c_{|M|}$ by $c_m$.
Then, by Lemma \ref{lemma:3},
\begin{align*}
    P\left( \max_\mathcal M | Z_M / \sqrt{c_m}| >1 \right) &\le \sum_{j=1}^{k m_0} \sum_{\mathcal M_j} P(Z_M^2 > c_j)\\
    &=  \sum_{j=1}^{k m_0} \binom{p}{j} P(\chi^2_1 > c_j)\le  \sum_{j=1}^{k m_0} \binom{p}{j} P(\chi^2_j > c_j) \to 0.
\end{align*}
Therefore, $|\bmu^T \left( I - \bH_M \right) \bepsilon| \le  \sqrt{\Delta_M}  |Z_M| \le \sqrt{\Delta_M} \sqrt{c_m} (1+o_p(1))$ uniformly over $\mathcal M$. Since $c_m = O(m_0 \log p)$, and by the identifiability condition \eq{idCon}, $m_0 \log p = o(\Delta_M)$ uniformly over $\mathcal M$ s.t. $M_0 \not\subset M$,
\[|\bmu^T \left( I - \bH_M \right) \bepsilon| = o_p(\Delta_M).\]

Now consider the third term in \eq{rssdiff}, by Lemma \ref{lemma:3} again,
\begin{align*}
    P\left( \max_\mathcal M \bepsilon^T \bH_M \bepsilon / c_m>1 \right) &\le \sum_{j=1}^{k m_0} \sum_{\mathcal M_j} P(\bepsilon^T \bH_M \bepsilon> c_j) =\sum_{j=1}^{k m_0} \binom{p}{j} P(\chi^2_j > c_j ) \to 0.
\end{align*}
So $\bepsilon^T \bH_M \bepsilon \le c_m (1+o_p(1))$ and
\[
\bepsilon^T \bH_M \bepsilon = o_p(\Delta_M)
\]
uniformly over $\mathcal M$ s.t. $M_0 \not\subset M$.

Therefore
\[
\text{RSS}_{M}-\text{RSS}_{M_0} =  \Delta(M) (1+o_p(1)),
\]
and
\[
\log\left( \frac{\text{RSS}_M}{\text{RSS}_{M_0}} \right) = \log \left( 1+\frac{\text{RSS}_M-\text{RSS}_{M_0}}{\text{RSS}_{M_0}} \right) = \log \left\{1+\frac{\Delta(M)}{n}(1+o_p(1))\right\}
\]
uniformly for all $M \in \mathcal M$ s.t. $M_0 \not\subset M$. Therefore
\begin{align*}
  T_1  &=  \frac{n(1+ o(1))}{2} \log\left\{ 1+ \frac{\Delta(M)}{n} (1+o_p(1))\right\} = \frac{\Delta(M) (1+ o_p(1))}{2}
  \end{align*}
uniformly for all $M \in \mathcal M$ s.t. $M_0 \not\subset M$.

Moreover,
\begin{align*}
&\frac{m-m_0}{2} \log(\pi \text{RSS}_{M_0})+\log  \left\{ \Gamma\left( \frac{n-m_0}{2} \right) / \Gamma\left( \frac{n-m}{2} \right) \right\} \\
=& \frac{m-m_0}{2} \log n (1+o_p(1)) +  \frac{m-m_0}{2} \log n (1+o(1))\\
=& (m-m_0)  \log n (1+o_p(1)).
\end{align*}

Finally,
\begin{align*}
    T_2 &= \frac{3}{2}  (m-m_0) \log n (1+o_p(1)) - \gamma \log \binom{p}{m_0} + \gamma \log \binom{p}{m} \\
&\ge -\frac{3}{2}m_0 \log n (1+o_p(1)) - \gamma m_0 \log p.
\end{align*}

\noindent {\bf Case 2: $M_0 \subset M$.}

Let $\mathcal M^* = \{M \in \mathcal M,\ M_0\subset M,\ M \ne M_0\}$ and $\mathcal M^*_j = \{M,\ |M| =j, M_0\subset M\}$.
First notice that $\text{RSS}_{M_0} - \text{RSS}_{M} = \chi^2_{m-m_0}(M)$, where $\chi^2_{m-m_0}(M)$ is a chi-square random variable depending on $M$ with degrees of freedom $m-m_0$.

 Recall $c_j = 2 j \left\{ \log p + \log (j \log p) \right\}$, by Lemma \ref{lemma:3} again,
\begin{align*}
    P\left(  \max_{1 \le j \le k m_0 - m_0} \max_{\mathcal M \in \mathcal M^*_j}  \chi^2_j(M)/ c_j  \ge 1\right) &\le \sum_{j=1}^{k m_0 - m_0} P\left(  \max_{\mathcal M \in \mathcal M^*_j}  \chi^2_j(M) \ge c_j \right)\\
    & =  \sum_{j=1}^{k m_0 - m_0}  \binom{p-m_0}{j} P\left(  \chi^2_j(M) \ge c_j \right) \\
    &\le \sum_{j=1}^{k m_0 - m_0}  \binom{p}{j} P\left(  \chi^2_j(M) \ge c_j \right)\to 0.
\end{align*}

It implies that
\[
\chi^2_{m-m_0}(M)\le c_{m-m_0} (1+o_p(1)).
\]

Note that $c_{m-m_0} = o(n)$ uniformly, therefore
\begin{align*}
    \frac{n-m-1}{2} \log\left( \frac{\text{RSS}_{M}}{\text{RSS}_{M_0}} \right) &= -\frac{n-m-1}{2}\log\left( 1+\frac{\chi^2_{m-m_0}(M)}{\text{RSS}_{M_0}-\chi^2_{m-m_0}(M)} \right) \\
   &\ge  -\frac{n-m-1}{2}\left( \frac{\chi^2_{m-m_0}(M)}{\text{RSS}_{M_0}-\chi^2_{m-m_0}(M)} \right) \\
    &\ge - \frac{c_{m-m_0}}{2} (1+o_p(1))\\
    &\ge -(m-m_0)\left[  1+ \frac{\log\{(k m_0-m_0) \log p\}  }{\log p}\right] \log p (1+o_p(1))\\
    &\ge  -(m-m_0)(1+ \delta) \log p (1+o_p(1))
\end{align*}
uniformly over $\mathcal M^*$.

Therefore, we show that
\[
T_1 \ge - (m-m_0) (1+\delta) \log p (1+o_p(1))
\]
uniformly over $\mathcal M^*$.

By Lemma \ref{lemma:1}, for $m_0<m<km_0$, $\log \binom{p}{m}  = (1-\delta) m \log p (1 + o(1))$
uniformly over $\mathcal M^*$.

Therefore,
\begin{align*}
T_2 =\frac{3}{2}(m-m_0) \log n (1+o_p(1))+ \gamma (1-\delta) (m-m_0) \log p (1 + o(1))\end{align*}
uniformly over $\mathcal M^*$.

Finally,
\begin{align*}
    \max_{M \ne M_0, M \in \mathcal M} r(M)/r(M_0) = \max\left\{ \max_{M_0 \not\subset M} \exp\left( -T_1 -T_2 \right), \max_{M_0 \subset M}\exp\left( -T_1 -T_2 \right) \right\}.
\end{align*}
By \eq{a1},
$\max_{M_0 \not\subset M}\exp\left( -T_1 -T_2 \right) \overset{P}{\to} 0$ since
\[
\min_{M_0 \not\subset M} T_1 + T_2 \to \infty
\]
and by \eq{a2},
$\max_{M_0 \subset M}\exp\left( -T_1 -T_2 \right) \to 0$  if  $\gamma> \frac{1+\delta}{1-\delta} - \frac{3\eta}{2(1-\delta)}$.
It proves that
\[
\max_{M \ne M_0, M \in \mathcal M} r(M)/r(M_0) \overset{P}{\to} 0.
\]
Moreover, if \eq{selectCon} holds,
\[
\sum_{M \ne M_0, M \in \mathcal M'} r(M)/r(M_0) \le \sum_{j=1}^{km_0} \sum_{\mathcal M'_j}  r(M)/r(M_0)  \le km_0  \max_{M \ne M_0, M \in \mathcal M} |M'_j| r(M)/r(M_0) \to 0.
\]
\bibliographystyle{rss}
\bibliography{Bibliography}

\end{document}